\documentclass[aps,prl,twocolumn,superscriptaddress,showpacs,showkeys,amsmath,amssymb]{revtex4-1}

\usepackage[T1]{fontenc}
\usepackage[utf8]{inputenc}
\usepackage{textcomp}
\usepackage{multirow}
\usepackage{float}
\usepackage{color}
\usepackage{bm}
\usepackage{amsmath}
\usepackage{amssymb}
\usepackage{graphicx}

\makeatletter

\usepackage[breaklinks,colorlinks,bookmarks=false,citecolor=blue,linkcolor=red,urlcolor=blue]{hyperref}
\renewcommand{\theequation}{\arabic{section}.\arabic{equation}}

\usepackage{latexsym}
\usepackage[ngerman,english]{babel}

\begin{document}

\title{Flat-band physics in the spin-1/2 sawtooth chain}

\author{Oleg~Derzhko}
\affiliation{Institute for Condensed Matter Physics,
          National Academy of Sciences of Ukraine,
          Svientsitskii Street 1, 79011 L'viv, Ukraine}
\affiliation{Max-Planck-Institut f\"{u}r Physik komplexer Systeme, 
          N\"{o}thnitzer Stra\ss e 38, 01187 Dresden, Germany}

\author{J\"{u}rgen~Schnack}          
\affiliation{Fakult\"{a}t f\"{u}r Physik, Universit\"{a}t Bielefeld,
          Postfach 100131, 33501 Bielefeld, Germany}

\author{D.~V.~Dmitriev}
\affiliation{Institute of Biochemical Physics of RAS,
          Kosygin Street 4, 119334 Moscow, Russia}

\author{V.~Ya.~Krivnov}
\affiliation{Institute of Biochemical Physics of RAS,
          Kosygin Street 4, 119334 Moscow, Russia}

\author{Johannes~Richter}
\affiliation{Institut f\"{u}r Physik,
          Otto-von-Guericke-Universit\"{a}t Magdeburg,
          P.O. Box 4120, 39016 Magdeburg, Germany}
\affiliation{Max-Planck-Institut f\"{u}r Physik komplexer Systeme, 
          N\"{o}thnitzer Stra\ss e 38, 01187 Dresden, Germany}
          
\date{\today}

\begin{abstract}
We consider the strongly anisotropic spin-1/2 $XXZ$ model on the sawtooth-chain lattice
with ferromagnetic longitudinal interaction $J^{zz}=\Delta J$ and aniferromagnetic transversal interaction $J^{xx}=J^{yy}=J>0$.
At $\Delta=-1/2$ the lowest one-magnon excitation band is dispersionless (flat) 
leading to a massively degenerate set of ground states.
Interestingly, this model admits a three-coloring representation of the ground-state manifold 
[H.~J.~Changlani et at., Phys. Rev. Lett. {\bf 120}, 117202 (2018)].
We characterize this ground-state manifold and elaborate the low-temperature thermodynamics of the system.
We illustrate the manifestation of the flat-band physics of the anisotropic model
by comparison with two isotropic flat-band Heisenberg sawtooth chains.
Our analytical consideration is complemented by exact diagonalization and finite-temperature Lanczos method calculations.
\end{abstract}



\maketitle

\section{Introduction}
\label{sec1}
\setcounter{equation}{0}

Frustrated quantum Heisenberg spin systems are of great interest nowadays.
Exact calculations and rigorous statements, although scarce, are obviously important for this field.
One source of such results stems from the flat-band antiferromagnets,
i.e., the models with a dispersionless (flat) one-magnon band \cite{Derzhko2015}.
The flat one-magnon band leads to localized multi-magnon states 
which dominate the low-temperature physics in antiferromagnetic flat-band models close to the saturation field. 
Their contribution to the partition function can be exactly calculated 
by visualizing the localized multi-magnon states as hard-core-object configurations on a corresponding auxiliary lattice.
Then the hard-core description allows to use classical statistical mechanics to describe frustrated quantum spin models.
This approach has been successfully used for a wide class of frustrated quantum antiferromagnets supporting flat bands
\cite{Zhitomirsky2004,Derzhko2004,Zhitomirsky2005,Derzhko2006,Zhitomirsky2007,Schnack2018} 
including the kagome antiferromagnet in two dimensions and the pyrochlore antiferromagnet in three dimensions. 
We mention that a similar description of flat-band states can be developed for the Hubbard model
\cite{Derzhko2015,Mielke1991,Tasaki1992,Derzhko2010,Maksymenko2012}. 
A popular one-dimensional example of a flat-band antiferromagnet is the Heisenberg sawtooth chain 
with the special relation between antiferromagnetic exchange interaction along the basal line $J_1>0$ and along the zig-zag path $J_2>0$,
of $J_2/J_1=2$, 
that was widely used as a playground for localized-magnon physics at low temperatures around the saturation field $h_{\rm sat}=4J_1$,  
see, e.g., \cite{Schulenburg2002,Zhitomirsky2004,Derzhko2004,Zhitomirsky2005,Derzhko2006,Richter2008,Metavitsiadis2020}.

Later on it was found 
that flat-band physics and corresponding localized multi-magnon states can appear in frustrated magnets also at zero magnetic field
in case that ferro- and antiferromagnetic interactions compete \cite{Krivnov2014}.
Again, the sawtooth chain is a prominent example, 
however, 
with ferromagnetic bonds $J_2<0$ along the zig-zag path and antiferromagnetic bonds $J_1>0$ along the basal line 
\cite{Tonegawa2004,Kaburagi2005,Krivnov2014,Dmitriev2015,Dmitriev2017,Dmitriev2019,Dmitriev2020}.
Here the flat-band physics is realized at a critical point $J_2/J_1=-2$, 
where the ferromagnetic ground state gives way for a ferrimagnetic one.  
It is worth mentioning 
that the ferro-antiferromagnetic sawtooth chain 
is an appropriate model to describe the recently synthesized compound Fe$_{10}$Gd$_{10}$ \cite{Baniodeh2018} 
and 
is also relevant for Cs$_2$LiTi$_3$F$_{12}$ that hosts ferro-antiferromagnetic sawtooth chains as magnetic subsystems \cite{Shirakami2019}.

Very recently, using the three-coloring description H.~J.~Changlani et al. \cite{Changlani2018,Changlani2019} have noticed
that the ground-state manifold of the spin-1/2 $XXZ$ sawtooth chain 
with antiferromagnetic bonds $J_1=J_2>0$ and with a negative $zz$ anisotropy parameter $\Delta=-1/2$ 
(denoted as $XXZ0$ model)
exhibits also a huge degeneracy. 
As already noticed before (but not investigated) in Ref.~\cite{Dmitriev2015}, 
the $XXZ0$ sawtooth chain also belongs to the class of flat-band systems hosting localized multi-magnon states in zero magnetic field.
The three-coloring description of spin systems is a general and promising approach to study frustrated magnets
\cite{Changlani2018,Changlani2019,Luban2001,Cepas2011,Jaubert2016}.
To illustrate the relation between the three-coloring and the flat-band localized-magnon description by the example of the sawtooth spin chain 
is one of the aims of the present study.

\begin{figure}
\begin{center}
\includegraphics[clip=on,width=80mm,angle=0]{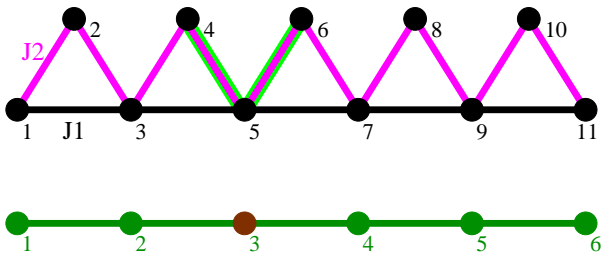}
\caption{(Top) The sawtooth-chain lattice 
(here, $N=11$ sites and open boundary conditions are imposed) 
considered in the present study. 
The greenly highlighted bonds mark a trap for a localized magnon on the sawtooth chain, see Eqs.~(\ref{205}) -- (\ref{208}).
(Bottom) Auxiliary simple chain used for characterization of the ground states of the sawtooth-chain spin models.
A localized magnon is represented by a brown site of the simple chain.}
\label{f01} 
\end{center}
\end{figure}

In the present paper,
we examine the spin-1/2 $XXZ0$ sawtooth-chain model \cite{Changlani2018}
focusing on the specific flat-band features, i.e., localized-magnon properties.
We also compare this model with the two isotropic Heisenberg sawtooth-chain flat-band cases 
which were mentioned above and studied previously 
\cite{Schulenburg2002,Zhitomirsky2004,Derzhko2004,Zhitomirsky2005,Derzhko2006,Richter2008,Krivnov2014,Dmitriev2015,Dmitriev2017,Dmitriev2019,Dmitriev2020}. 
To be specific, in what follows we consider the spin-1/2 $XXZ$ Hamiltonian
\begin{eqnarray}
\label{101}
H
=
J_1\sum_{i=1}^{{\cal {N}}}\left(s_{2i-1}^x s_{2i+1}^x + s_{2i-1}^y s_{2i+1}^y +\Delta_1 s_{2i-1}^z s_{2i+1}^z\right)
\nonumber\\
+
J_2\sum_{i=1}^{{\cal {N}}}\left(s_{2i-1}^x s_{2i}^x + s_{2i-1}^y s_{2i}^y +\Delta_2 s_{2i-1}^z s_{2i}^z
\right.  
\hspace*{1cm}
\nonumber\\
\left.
+s_{2i}^x s_{2i+1}^x + s_{2i}^y s_{2i+1}^y +\Delta_2 s_{2i}^z s_{2i+1}^z
\right)
-h\sum_{i=1}^Ns_i^z 
\qquad
\end{eqnarray}
on the sawtooth-chain lattice of $N$ sites (see Fig.~\ref{f01}),
where we have ${\cal{N}}=(N-1)/2$ for open boundary conditions and ${\cal{N}}=N/2$ for periodic boundary conditions.
In what follows we choose the following flat-band parameter sets: 
\begin{itemize}
\item model 1: $J_2=J_1>0$, $\Delta_1=\Delta_2=-1/2$ \cite{Changlani2018};
\item model 2: $J_2=-2J_1<0$, $\Delta_1=\Delta_2=1$ \cite{Krivnov2014};
\item model 3: $J_2=2J_1>0$, $\Delta_1=\Delta_2=1$ \cite{Derzhko2006},
\end{itemize}
where model 1 corresponds to the $XXZ0$ model mentioned above.
It is convenient to set $J_1=1$ for model 1 \cite{Changlani2018}, but $J_1=1/2$ for models 2 \cite{Krivnov2014} and 3.

In addition to analytical investigations of the models presented in Secs.~\ref{sec2} -- \ref{sec4}
we will use 
full exact diagonalization (ED) employing J.~Schulenburg's \textit{spinpack} code \cite{spinpack} 
and 
the finite-temperature Lanczos (FTL) technique \cite{FTL1,FTL2,FTL3,FTL4} 
to discuss numerical data for finite sawtooth chains in Secs.~\ref{sec4} and \ref{sec5}.

\section{Constituents for many-body physics}
\label{sec2}
\setcounter{equation}{0}

We begin with the illustration of some key elements 
relevant for the localized-magnon picture and for the three-coloring representation.
First of all we note that the spin Hamiltonian $H$ commutes with total $S^z=\sum_{i=1}^Ns_i^z$ 
that allows us to consider the eigenstates of the Hamiltonian in each of $N+1$ subspaces of $S^z=N/2,N/2-1,\ldots,-N/2$ separately.
Clearly, 
the fully polarized ferromagnetic state $\vert 0\rangle$ is the only eigenstate of $H$ in the subspace with $S^z=N/2$ with the energy $E_{0}$ 
and it can be considered as the magnon vacuum state. 
It is straightforward to get the eigenstates and eigenvalues in the subspace with $S^z=N/2-1$ (one-magnon states), 
see below.

Since the sawtooth chain is a one-dimensional array of corner-sharing triangles,
see Fig.~\ref{f01},
its Hamiltonian can be written as a sum over Hamiltonians of each triangle,
\begin{eqnarray}
\label{201}
H&=&\sum_{\triangle}H_{\triangle},
\nonumber\\
H_{\triangle}
&=&
J_1\left(s_{2i-1}^x s_{2i+1}^x + s_{2i-1}^y s_{2i+1}^y +\Delta_1 s_{2i-1}^z s_{2i+1}^z\right)
\nonumber\\
&&+
J_2\left(s_{2i-1}^x s_{2i}^x + s_{2i-1}^y s_{2i}^y +\Delta_2 s_{2i-1}^z s_{2i}^z
\right.
\nonumber\\
&&\left.
\qquad +s_{2i}^x s_{2i+1}^x + s_{2i}^y s_{2i+1}^y +\Delta_2 s_{2i}^z s_{2i+1}^z\right)
\end{eqnarray}
(the Zeeman term is omitted).
Below we also discuss the eigenstates and eigenvalues of the spin Hamiltonian $H_\triangle$ 
as they 
explain the three-coloring representation for the sawtooth-chain spin model 1 
and
provide a route to construct the eigenstates of the sawtooth-chain spin models 1 and 2.

\subsection{Flat bands and localized magnons}
\label{sec2a}

Imposing periodic boundary conditions, 
$s_{N+1}^\alpha=s_1^\alpha$ ($N$ is even),
we find straightforwardly the energies of the one-magnon excitations above the fully polarized ferromagnetic state $\vert 0\rangle$ 
(magnon vacuum)
for all three spin models.
All of them exhibit flat bands.
For the models at hand we have
\begin{eqnarray}
\label{202}
E_1(k)-E_0=0,  
\;\;\; 
E_2(k)-E_0= \frac{3}{2}+\cos k
\end{eqnarray}
for model 1,
\begin{eqnarray}
\label{203}
E_1(k)-E_0=0,  
\;\;\; 
E_2(k)-E_0= \frac{3}{2}+\frac{1}{2}\cos k
\end{eqnarray}
for model 2,
and
\begin{eqnarray}
\label{204}
E_1(k)-E_0=-2, 
\;\;\; 
E_2(k)-E_0= -\frac{1}{2}+\frac{1}{2}\cos k 
\end{eqnarray}
for model 3 
(recall that $J_1=1/2$ for models 2 and 3), 
where $E_0$ is the energy of the ferromagnetic state.
Here, as usually, $k$ acquires $N/2$ values within the region between $-\pi$ and $\pi$.
Note that in the cases 1 and 2 the (lowest-energy) flat-band excitations have zero energy.

Furthermore,
one can construct the flat-band states as localized states where a magnon is located on three adjacent sites of the lattice (magnon trap), 
see Fig.~\ref{f01}, where a trap built by sites $4,5,6$ is greenly highlighted.
We have
\begin{eqnarray}
\label{205}
\vert l_i\rangle=l_i\vert 0\rangle,
\;\;\;
i=1,\ldots,\frac{N}{2}
\end{eqnarray}
with
\begin{eqnarray}
\label{206}
l_i=s_{2i-2}^--s_{2i-1}^-+s_{2i}^-
\end{eqnarray}
for model 1,
\begin{eqnarray}
\label{207}
l_i=s_{2i-2}^-+2s_{2i-1}^-+s_{2i}^-
\end{eqnarray}
for model 2,
and
\begin{eqnarray}
\label{208}
l_i=s_{2i-2}^--2s_{2i-1}^-+s_{2i}^-
\end{eqnarray}
for model 3,
see Fig.~\ref{f01}.
Note that with periodic boundary conditions $0\equiv N$, i.e., $s_{0}^{-} \equiv s_{N}^{-}$.
The local nature of the one-magnon ground states (\ref{205}) allows to construct many-magnon ground states, see Sec.~\ref{sec3}.

\subsection{Spin model on a triangle}
\label{sec2b}

Here we provide some formulas which we need in the following sections. 
The eigenstates of $H_\triangle$ (\ref{201}) for the model 1 may be written in the form
\begin{eqnarray}
\label{209}
\vert 1\rangle&=&\vert\uparrow\uparrow\uparrow\rangle,
\nonumber\\
\vert 2_{\chi}\rangle&=&
\vert\downarrow\uparrow\uparrow\rangle+\omega \vert\uparrow\downarrow\uparrow\rangle+\omega^2 \vert\uparrow\uparrow\downarrow\rangle,
\;\;\;
\omega=\exp\frac{2\pi{\rm{i}}}{3},
\nonumber\\
\vert 3_{\chi}\rangle&=&
\vert\downarrow\uparrow\uparrow\rangle+\omega^2 \vert\uparrow\downarrow\uparrow\rangle+\omega \vert\uparrow\uparrow\downarrow\rangle,
\nonumber\\
\vert 5_{\chi}\rangle&=&
\vert\uparrow\downarrow\downarrow\rangle+\omega \vert\downarrow\uparrow\downarrow\rangle+\omega^2 \vert\downarrow\downarrow\uparrow\rangle,
\nonumber\\
\vert 6_{\chi}\rangle&=&
\vert\uparrow\downarrow\downarrow\rangle+\omega^2 \vert\downarrow\uparrow\downarrow\rangle+\omega \vert\downarrow\downarrow\uparrow\rangle,
\nonumber\\
\vert 8\rangle&=&\vert\downarrow\downarrow\downarrow\rangle
\end{eqnarray}
(all of them have the eigenvalue $-3/8$)
and 
\begin{eqnarray}
\label{210}
\vert 4\rangle&=&
\vert\downarrow\uparrow\uparrow\rangle+\vert\uparrow\downarrow\uparrow\rangle+ \vert\uparrow\uparrow\downarrow\rangle,
\nonumber\\
\vert 7\rangle&=&
\vert\uparrow\downarrow\downarrow\rangle+ \vert\downarrow\uparrow\downarrow\rangle+ \vert\downarrow\downarrow\uparrow\rangle
\end{eqnarray}
(both have the eigenvalue $9/8$).
The eigenstates 
$\vert 2_{\chi}\rangle$, 
$\vert 3_{\chi}\rangle$, 
$\vert 5_{\chi}\rangle$, 
and
$\vert 6_{\chi}\rangle$, 
are also the eigenstates of the chirality operator of the triangle
and thus their form is important for constructing a three-coloring representation for the ground-state manifold of model 1 \cite{Changlani2018}.
However, to get a better relation to the investigations of Refs.~\cite{Krivnov2014,Dmitriev2015} 
we may introduce other linear combinations of these states,
namely,
\begin{eqnarray}
\label{211}
\vert 2\rangle=\vert\downarrow\uparrow\uparrow\rangle - \vert\uparrow\downarrow\uparrow\rangle
\propto \vert 2_{\chi}\rangle-\omega \vert 3_{\chi}\rangle,
\nonumber\\
\vert 3\rangle=\vert\downarrow\uparrow\uparrow\rangle - \vert\uparrow\uparrow\downarrow\rangle
\propto \omega\vert 2_{\chi}\rangle- \vert 3_{\chi}\rangle,
\nonumber\\
\vert 5\rangle=\vert\uparrow\downarrow\downarrow\rangle - \vert\downarrow\uparrow\downarrow\rangle
\propto \vert 5_{\chi}\rangle-\omega \vert 6_{\chi}\rangle,
\nonumber\\
\vert 6\rangle=\vert\uparrow\downarrow\downarrow\rangle - \vert\downarrow\downarrow\uparrow\rangle
\propto \omega\vert 5_{\chi}\rangle- \vert 6_{\chi}\rangle.
\end{eqnarray}
We use these eigenstates in Sec.~\ref{sec3} while constructing many-magnon ground states.

For the model 2, 
the states  
$\vert\!\uparrow\uparrow\uparrow\rangle$,
$\vert\downarrow\uparrow\uparrow\rangle+ \vert\uparrow\downarrow\uparrow\rangle+ \vert\uparrow\uparrow\downarrow\rangle$,
$\vert\downarrow\uparrow\uparrow\rangle-\vert\uparrow\uparrow\downarrow\rangle$, 
$\vert\uparrow\downarrow\downarrow\rangle+ \vert\downarrow\uparrow\downarrow\rangle+ \vert\downarrow\downarrow\uparrow\rangle$,
$\vert\uparrow\downarrow\downarrow\rangle-\vert\downarrow\downarrow\uparrow\rangle$,
$\vert\downarrow\downarrow\downarrow\rangle$
are the eigenstates with the eigenvalue $-3/8$
and 
the states 
$\vert\downarrow\uparrow\uparrow\rangle-2\vert\uparrow\downarrow\uparrow\rangle+ \vert\uparrow\uparrow\downarrow\rangle$,
$\vert\uparrow\downarrow\downarrow\rangle-2 \vert\downarrow\uparrow\downarrow\rangle+ \vert\downarrow\downarrow\uparrow\rangle$
are the eigenstates with the eigenvalue $9/8$. 

For consistency, we give also the states for the model 3.
This set of states has a completely different structure:
The highest-energy is the quadruplet  
($\vert\!\uparrow\uparrow\uparrow\rangle$,
$\vert\!\downarrow\uparrow\uparrow\rangle + \vert\!\uparrow\downarrow\uparrow\rangle + \vert\!\uparrow\uparrow\downarrow\rangle$ 
etc.) 
with the energy $5/8$,
the two states 
$\vert\downarrow\uparrow\uparrow\rangle-\vert\uparrow\uparrow\downarrow\rangle$
and
$\vert\uparrow\downarrow\downarrow\rangle-\vert\downarrow\downarrow\uparrow\rangle$
have the energy $-3/8$,
and finally
the two states 
$\vert\downarrow\uparrow\uparrow\rangle-2\vert\uparrow\downarrow\uparrow\rangle +\vert\uparrow\uparrow\downarrow\rangle$
and
$\vert\uparrow\downarrow\downarrow\rangle -2\vert\downarrow\uparrow\downarrow\rangle +\vert\downarrow\downarrow\uparrow\rangle$
have the energy $-7/8$.

\section{Many-magnon states}
\label{sec3}
\setcounter{equation}{0}

So far only for the models 2 and 3 the construction of localized-magnon states was described in the literature, 
see, e.g., Ref.~\cite{Krivnov2014} for model 2 and Refs.~\cite{Schulenburg2002,Zhitomirsky2004,Derzhko2004} for model 3,
but not for model 1.
Therefore, we sketch now the construction rules of localized-magnon states for model 1 in this section. 
In addition, we will briefly discuss the relation of the localized-magnon states to the promising three-coloring picture 
developed in Refs.~\cite{Changlani2018,Changlani2019}.

We begin with a short outline of the three-coloring representation for the ground-state manifold of the spin-1/2 $XXZ0$ sawtooth chain
(i.e., model 1 in our notation).
The starting point is the definition of three single-spin coloring states
\begin{equation}
\label{301}
\vert r\rangle\equiv\vert\uparrow\rangle + \vert\downarrow\rangle,
\vert b\rangle\equiv\vert\uparrow\rangle + \omega\vert\downarrow\rangle,
\vert g\rangle\equiv\vert\uparrow\rangle + \omega^2\vert\downarrow\rangle,
\end{equation}
where $\omega=\exp(2\pi{\rm{i}}/3)$, see Eq.~(\ref{209}).
These states are represented by the colors red, blue, green, respectively.
A multi-spin state on a lattice is constructed by putting a single-spin coloring state at each lattice site. 
The state can be graphically represented as a three-coloring of the lattice
(i.e., no two vertices connected by a bond have the same color).
Obviously, the two three-coloring states
$\vert r_1b_2g_3\rangle$ and $\vert r_1g_2b_3\rangle$ on a triangle 
are superpositions of states given in Eq.~(\ref{209}) with the ground-state energy $-3/8$,
where states with different $S^z$ are mixed.
In other words,
the states $\vert r_1b_2g_3\rangle$ and $\vert r_1g_2b_3\rangle$ belong to the ground-state manifold.
The three-coloring can be straightforwardly extended to a thermodynamically large lattice.
The total number of three-colorings for the open sawtooth chain of $N=2{\cal{N}}+1$ sites grows as $2^{\cal{N}}$,
where ${\cal{N}}=(N-1)/2$ is the number of triangles in the open sawtooth chain.
Moreover,
constructing resonating color loops, one can single out a localized magnon state \cite{Changlani2019},
see also below for an example.
However, by contrast to the flat-band localized-magnon description,
the utilization of the three-coloring picture
to determine properties of corresponding frustrated spin models, such as model 1, is much less elaborated, 
i.e., it is still a task to be addressed in the future. 
One difficulty is certainly the mixing of states with different $S^z$ 
that requires a subsequent projection onto the $S^z$-subspaces to restore this symmetry of Hamiltonian (\ref{101}).   
In what follows, 
we therefore mainly exploit the localized-magnon picture for the one-dimensional sawtooth-chain spin model.

First we mention that in the subspace $S^z=N/2-1$ 
the localized-magnon (or flat-band) states introduced in Sec.~\ref{sec2} are exact eigenstates. 
For the sawtooth chain model 1 of $N=2{\cal N}+1$ sites with open boundary conditions
there are two classes of localized one-magnon states, 
namely, 
``boundary'' states such as
\begin{eqnarray}
\label{302}
\vert l_1\rangle
= 
\left(-\vert\downarrow_1\uparrow_2\uparrow_3\rangle + \vert\uparrow_1\downarrow_2\uparrow_3\rangle\right)\vert\ldots\uparrow\ldots\rangle
\end{eqnarray}
and 
``bulk'' states such as
\begin{eqnarray}
\label{303}
\vert l_2\rangle
= 
\left(\vert\downarrow_2\uparrow_3\uparrow_4\rangle-\vert\uparrow_2\downarrow_3\uparrow_4\rangle + \vert\uparrow_2\uparrow_3\downarrow_4\rangle\right)
\vert\ldots\uparrow\ldots\rangle , \qquad
\end{eqnarray}
where the numbers at the up- and down-arrows correspond to the numbering in Fig.~\ref{f01}, top.
Both states belong to the ground-state manifold.
(For an explicit proof we refer to Appendix~\ref{appendix}.)
Note here that the localized boundary ground states exist also for model 2 \cite{footnote1} but not for model 3.

Let us also give an example how to get a localized magnon state from the three-coloring representation.
We have
\begin{eqnarray}
\label{304}
\!\!\!\!\!\!\!
\vert l_2\rangle
\propto
P_{S^z=\frac{N}{2}-1}
\!\left(\vert r_1b_2g_3b_4r_5\ldots \rangle \!-\! \vert r_1g_2b_3g_4r_5\ldots \rangle \right)\!,
\end{eqnarray}
where $P_{S^z}$ stands for the projector onto the subspace with the specific $S^z$
and the numbers $1,\ldots,5$ correspond to those given in the first line of Fig.~\ref{f01}.

In summary,
the ground-state degeneracy for the open sawtooth chain 1 of $N=2{\cal{N}}+1$ sites in the subspace $S^z=N/2-1$ is ${\cal{N}}+1$, 
because all the localized-magnon states are linearly independent \cite{Schmidt2006}.  
A corresponding consideration holds for the open sawtooth-chain model 2,
i.e., the ground-state degeneracy is also ${\cal {N}}+1$.
On the other hand, 
for the open sawtooth-chain model 3 the degeneracy in the subspace $S^z=N/2-1$ is lower and equals ${\cal {N}}-1$, 
because the localized boundary states are missing.

\begin{figure}
\begin{center}
\includegraphics[clip=on,width=80mm,angle=0]{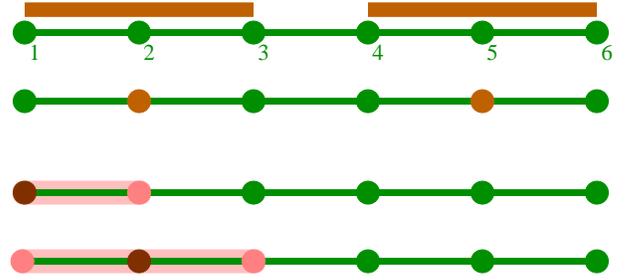}
\caption{Visualization of the ground states in the subspace $S^z=N/2-2$ on the auxiliary linear chain (open boundary conditions) 
which corresponds to the  $N=11$ sawtooth chain.
Line 1: 
Two independent localized magnons can be pictorially represented as a spatial configuration of two hard dimers.
A hard dimer extends over two lattice constants. 
Overlapping of two dimers is forbidden, 
i.e., two neighboring sites of the auxiliary linear chain cannot be occupied by independent localized magnons.
Line 2: 
Corresponding position of the localized magnons (filled brown circles).   
Lines 3 and 4: 
Two different two-magnon complexes of overlapping localized magnons 
corresponding to Eq.~(\ref{305}) (line 3) and to Eq.~(\ref{306}) (line 4).}
\label{f02} 
\end{center}
\end{figure}

We pass to the subspace $S^z=N/2-k$ with $k=2$ down spins.
Because of the localized nature of the one-magnon excitations,
independent localized two-magnon eigenstates can be constructed satisfying the hard-dimer rule (see lines 1 and 2 in Fig.~\ref{f02}),
i.e., two localized one-magnon states are not allowed to be in touch. 
There are ${\cal C}_{{\cal N}-k+2}^k$, $k=2$, such states for the open sawtooth chain of $N=2{\cal N}+1$ sites,
where ${\cal C}_{m}^{n}=m!/[n!(m-n)!]$ is the binomial coefficient.
This construction rule was first found for model 3 and can be extended to more than two magnons
(so called independent localized multi-magnon states) 
leading finally to a huge ground-state degeneracy of model 3 at the saturation field $h_{\rm sat}$ 
that grows exponentially with system size $N$, cf. \cite{Schulenburg2002,Zhitomirsky2004,Derzhko2004}.
Obviously, 
the above illustrated construction of independent localized multi-magnon states also holds for models 1 and 2 \cite{Krivnov2014,Dmitriev2015};
the (natural) number of the magnons $k$ for these open chains varies in the region $2\le k\le({\cal N}+2)/2$. 
However, there are two important differences to model 3:
(i) all these multi-magnon states are degenerate at zero field $h=0$ [cf. Eqs.~(\ref{202}), (\ref{203}) and Eq.~(\ref{204})]   
and 
(ii) in addition to the independent localized-magnon states 
also specifically overlapping localized magnons are ground states \cite{Krivnov2014}.
Thus, the ground-state degeneracy of models 1 and 2 is much larger than for model 3.

To illustrate  overlapping localized magnons, 
we consider a localized two-magnon complex at the boundary defined as \cite{Krivnov2014,Dmitriev2015}
\begin{eqnarray}
\label{305}
l_1\left(cl_{1} + l_{2}\right)\vert 0\rangle
=
l_1\left(- \frac{l_{1}}{2} + l_{2}\right)\vert 0\rangle  \quad &&
\nonumber\\
=
\left(-s_1^-+s_2^-\right)\left(\frac{s_1^-}{2}+\frac{s_2^-}{2}-s_3^-+s_4^-\right)\vert 0\rangle
&&
\end{eqnarray}
($c=-1/2$),
see line 3 in Fig.~\ref{f02} for a pictorial representation of the state \eqref{305}. 
In Appendix~\ref{appendix} we check that this state is among the ground states with $S^z=N/2-2$.

A two-magnon complex away from the boundary is given by the formula
\begin{eqnarray}
\label{306}
l_2\left(l_{1}+cl_{2} + l_{3}\right)\vert 0\rangle
=
l_2\left(l_{1}- \frac{l_{2}}{2} + l_{3}\right)\vert 0\rangle
\nonumber\\
=\left(\!s_2^-\!-\!s_3^-\!+\!s_4^-\!\right)
\!\left(\!-s_1^-\!+\!\frac{s_2^-}{2}\!+\!\frac{s_3^-}{2}\!+\!\frac{s_4^-}{2}\!-\!s_5^-\!+\!s_6^-\!\right)\!\vert 0\rangle
\qquad
\end{eqnarray}
($c=-1/2$),
see line 4 in Fig.~\ref{f02}. 
Again, in Appendix~\ref{appendix} we check that this state is among the ground states with $S^z=N/2-2$.

It is easy to count the ground states in the sector $S^z=N/2-2$.
We have 
${\cal{C}}_{\cal {N}}^{2}$ independent localized two-magnon states
and
${\cal N}+1$ localized  states built by localized two-magnon complexes,
in total
\begin{eqnarray}
\label{307}
g_{\cal N}(S^z=N/2-2)={\cal{C}}_{\cal {N}}^{2}+{\cal N}+1
=\sum_{k=0}^2{\cal{C}}_{\cal {N}}^{k}.
\end{eqnarray}
We have checked Eq.~(\ref{307}) by exact diagonalization of open sawtooth chains 1 of up to $N=39$ sites 
providing evidence for the completeness of the constructed ground states in the subspace $S^z=N/2-2$.
Moreover, these states are linearly independent \cite{Schmidt2006}.

\begin{figure}
\begin{center}
\includegraphics[clip=on,width=80mm,angle=0]{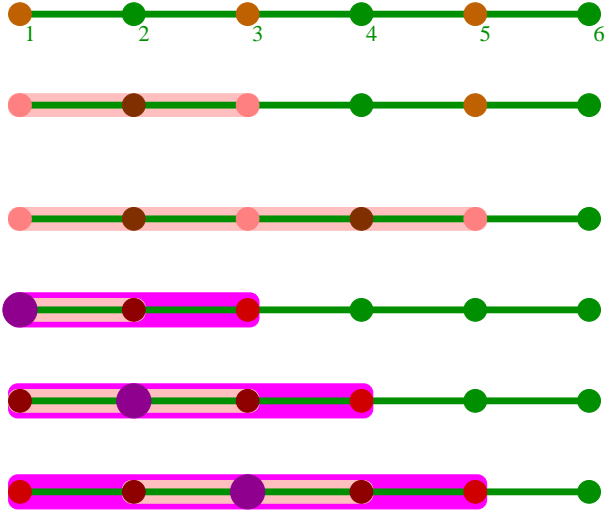}
\caption{Visualization of the ground states in the subspace $S^z=N/2-3$
(here $N=11$ sites, open boundary conditions).
Line 1: Three independent localized magnons can be pictorially represented as a spatial configuration of three hard dimers.
Line 2: Localized  magnon and localized two-magnon complex.
Line 3: Localized three-magnon complex of one-bracket type, see Eq.~\eqref{308}.
Lines 4, 5, and 6: Localized three-magnon complexes of two-bracket type, see Eq.~\eqref{310}.}
\label{f03} 
\end{center}
\end{figure}

Now we turn to the subspace $S^z=N/2-k$ with $k=3$ spins down.
Again, because of the local nature of the independent localized-magnon states and the localized two-magnon complexes,
we can construct a number of ground states in the sector $S^z=N/2-3$ placing such states sufficiently far from each other.
This way we construct 
${\cal{C}}_{{\cal N}-1}^{3}$ independent localized magnon states 
and 
$2{\cal{C}}_{{\cal N}-1}^{2}$ states consisting of a localized magnon and a localized two-magnon complex, 
see lines 1 and 2 in Fig.~\ref{f03} for $N=11$.
More ground states in this subspace are the localized three-magnon complexes,
where we have two types, 
which we will denote as one-bracket type and two-bracket type, see Eqs.~(\ref{308}), (\ref{309}) and Eq.~(\ref{310}) below.
These localized three-magnon complexes are sketched in Fig.~\ref{f03} for $N=11$,  
see line 3 for the one-bracket type  and lines 4, 5, and 6 for the two-bracket type.

An example of a localized three-magnon complex of one-bracket type that belongs to the ground-state manifold with $S^z=N/2-3$,  
see Appendix~\ref{appendix}, 
is given by 
\begin{eqnarray}
\label{308}
l_2l_4\left(l_{1}+cl_{2} + l_{3} +cl_{4} + l_{5}\right)\vert 0\rangle
\end{eqnarray}
($c=-1/2$),
see line 3 in Fig.~\ref{f03}.
Other three-magnon complexes of one-bracket type are given by the formulas:
\begin{eqnarray}
\label{309}
l_1l_3\left(cl_{1} + l_{2} +cl_{3} + l_{4}\right)\vert 0\rangle,
\nonumber\\
l_3l_5\left(l_{2}+cl_{3} + l_{4} +cl_{5} + l_{6}\right)\vert 0\rangle,
\nonumber\\
\vdots
\nonumber\\
l_{{\cal{N}}-1}l_{{\cal{N}}+1}\left(l_{{\cal{N}}-2}+cl_{{\cal{N}}-1} + l_{{\cal{N}}} +cl_{{\cal{N}}+1}\right)\vert 0\rangle.
\end{eqnarray}
Altogether, there are ${\cal{N}}-1$ three-magnon complexes of one-bracket type.

We pass to the ${\cal{N}}+1$  localized three-magnon complexes of two-bracket type
\begin{eqnarray}
\label{310}
l_1
\!\left(cl_1\!+\!l_2\right)\!\left(c^2l_{1}\!+\!cl_{2}\!+\!l_{3}\right)\!\vert 0\rangle
+d l_1^3\vert 0\rangle, \qquad
\nonumber\\
l_2
\!\left(l_1\!+\!cl_2\!+\!l_3\right)\!\left(cl_1\!+\!c^2l_{2}\!+\!cl_{3}\!+\!l_{4}\right)\!\vert 0\rangle
+d l_2^3\vert 0\rangle, \qquad
\nonumber\\
l_3
\!\left(l_2\!+\!cl_3\!+\!l_4\right)\!\left(l_1\!+\!cl_2\!+\!c^2l_{3}\!+\!cl_{4}\!+\!l_{5}\right)\!\vert 0\rangle
+d l_3^3\vert 0\rangle,  \qquad
\nonumber\\
\vdots \quad \qquad
\nonumber\\
l_{{\cal{N}}\!+\!1}
\!\left(l_{{\cal{N}}}\!+\!cl_{{\cal{N}}\!+\!1}\right)\!\left(l_{{\cal{N}}\!-\!1}\!+\!cl_{{\cal{N}}}\!+\!c^2l_{{\cal{N}}\!+\!1}\right)\!\vert 0\rangle
+d l_{{\cal{N}}\!+\!1}^3\vert 0\rangle  \qquad
\end{eqnarray}
($c=-1/2$ and $d=1/8$),
see lines 4, 5, and 6 in Fig.~\ref{f03}.
(Note that the last term in the first and the last lines of Eq.~(\ref{310}) is redundant
[since $l_1^3=(-s^-_1+s^-_2)^3=0$ and $l_{{\cal{N}}+1}^3=(s^-_{{\cal{N}}}-s^-_{{\cal{N}}+1})^3=0$],
it is written for similarity to the other lines, where this kind of terms are relevant.)  
Again, some more detailed calculations checking that the above presented states are ground states with $S^z=N/2-3$ 
are transferred to Appendix~\ref{appendix}. 

In sum, the number of ground states in the sector $S^z=N/2-3$ is
\begin{eqnarray}
\label{311}
g_{\cal N}(S^z\!=\!N/2\!-\!3)
&=&{\cal{C}}_{{\cal {N}}-1}^{3}+2{\cal{C}}_{{\cal {N}}-1}^{2} +{{\cal N}}\!-\!1+{{\cal N}}\!+\!1
\nonumber\\
&=&\sum_{k=0}^3{\cal{C}}_{\cal {N}}^{k}.
\end{eqnarray}
As previously, we have confirmed the analytical expression (\ref{311})
by exact diagonalization for models 1 with open boundary conditions of up to $N=39$ sites.
Moreover, these states are linearly independent.

The construction of the linearly independent ground states for $k\ge 4$ follows the same lines as explained above, 
although it becomes more tedious.
We find that model 1 with open boundary conditions is identical to model 2 \cite{Krivnov2014}.
After all, 
the degeneracy of the ground-state manifold of the open sawtooth-chain model 1 with $N=2{\cal{N}}+1$ sites 
in the subspaces $S^z=N/2-k$ for $k=0,1,2,\ldots,{\cal{N}}$
is given by
\begin{eqnarray}
\label{312}
g_{\cal{N}}(S^z=N/2-k) = {\cal{C}}_{\cal{N}}^0+{\cal{C}}_{\cal{N}}^1+\ldots +{\cal{C}}_{\cal{N}}^k,
\end{eqnarray} 
cf. Eqs.~\eqref{307} and \eqref{311}.
Then the total degeneracy of the ground-state manifold of the open sawtooth-chain model 1 with $N=2{\cal{N}}+1$ sites is
\begin{eqnarray}
\label{313}
{\cal{W}}(N) 
\!=\!2\sum_{k=0}^{\cal{N}}\left({\cal{C}}_{\cal{N}}^0\!+\!{\cal{C}}_{\cal{N}}^1\!+\ldots+\!{\cal{C}}_{\cal{N}}^k\right)
\!=\!2\sum_{k=0}^{\cal{N}}\left({\cal{N}}\!+\!1\!-\!k\right){\cal{C}}_{\cal{N}}^k
\nonumber\\
=2\left[\left({\cal{N}}+1\right)2^{\cal{N}} - {\cal{N}}2^{{\cal{N}}-1}\right]
=\left({\cal{N}}+2\right)2^{\cal{N}}.
\end{eqnarray}
As already found for the sectors with $k=2$ and $3$, 
the general formula \eqref{312} and of course also \eqref{313} match perfectly with corresponding exact-diagonalization data.

We add here the known information on the degeneracy of the ground-state manifold in the subspaces $S^z=N/2-k$ for $k=0,1,2,\ldots,{\cal{N}}$ 
of the models 2 and 3 with $N=2{\cal{N}}+1$ sites and open boundary conditions.
For model 2 Eq.~(\ref{312}) holds \cite{Krivnov2014}
and 
for model 3 the flat-band states exist only in the subspaces $S^z=N/2-k$, $k=0,1,2,\ldots,{\cal{N}}/2$, 
and the degeneracy is $g_{\cal{N}}(S^z=N/2-k)={\cal{C}}_{{\cal{N}}-k}^k$, 
i.e., it is smaller, because only independent localized multi-magnon states exist, but no additional complexes
\cite{Zhitomirsky2004,Derzhko2004,Derzhko2006}.

Let us now briefly discuss the ground-state degeneracy of periodic sawtooth chains of $N=2{\cal N}$ sites.
For model 3 independent localized multi-magnon ground states exist in the subspace $S^z=N/2-k$ with $k=0,1,2,\ldots,{\cal{N}}/2$. 
Their degeneracy is 
$G^{(3)}_{\cal N}(S^z)=[{\cal N}/({\cal N}-k)]{\cal C}_{{\cal N}-k}^k$
\cite{Zhitomirsky2004,Derzhko2004,Derzhko2006}.

For model 2,
the ground states in the subspace $S^z=N/2-k$ with $k=0,1,2,\ldots,{\cal{N}}$ were found in Refs.~\cite{Krivnov2014,Dmitriev2015};
their degeneracy is 
\begin{eqnarray}
\label{314}
G^{(2)}_{\cal N}(S^z)
=
\left\{
\begin{array}{ll}
{\cal C}_{{\cal N}}^k,                       & k=0,1,\ldots,\frac{{\cal{N}}}{2}, \\
{\cal C}_{{\cal N}}^{\frac{{\cal{N}}}{2}},   & k=\frac{{\cal{N}}}{2},\ldots,{\cal{N}}-1,\\
{\cal C}_{{\cal N}}^{\frac{{\cal{N}}}{2}}+1, & k={\cal{N}}.
\end{array}
\right.
\end{eqnarray}

For periodic chains, the model 1 exhibits more ground states than the model 2 if $k\ge 3$:
\begin{eqnarray}
\label{315}
&&
G^{(1)}_{\cal N}(S^z)=G^{(2)}_{\cal N}(S^z)+G^{\rm add}_{\cal N}(S^z),
\\
&&
G^{\rm add}_{\cal N}(S^z)
\!=\!
\left\{
\begin{array}{ll}
{\cal C}_{\cal N}^{k-3},                                                        & k=3,\ldots,\frac{{\cal N}}{2}, \\
2{\cal C}_{\cal N}^{\frac{{\cal N}}{2}-3}\!-\!{\cal C}_{\cal N}^{{\cal N}-k-3}, & k=\frac{{\cal N}}{2}\!+\!1,\ldots,{\cal N}\!-\!3, \\
2{\cal C}_{\cal N}^{\frac{{\cal N}}{2}-3},                                      & k={\cal N}\!-\!2,{\cal N}\!-\!1,{\cal N}.
\end{array}
\right.
\nonumber
\end{eqnarray}
The total degeneracy of the ground-state manifold of the periodic sawtooth-chain model 1 with $N=2{\cal{N}}$ sites then is
\begin{equation}
\label{316}
{\cal W}(N)=\left(\frac{\cal N}{3}+1\right)2^{\cal N} + \frac{2 {\cal N}}{3} + 1, 
\end{equation}
cf. Eq.~(\ref{313}).
We confirmed the numbers given in Eqs.~\eqref{314}, \eqref{315} and \eqref{316} 
by exact diagonalization for the periodic sawtooth-chain model 1 of up to $N=32$ sites
(see also Table~\ref{tab1}). 

Apparently, the ground-state degeneracies depend on the imposed boundary conditions for finite chains,
but in the thermodynamic limit $N\to\infty$ the boundary conditions become irrelevant.
Therefore, it is sufficient to consider for the analytical calculations of low-temperature thermodynamic quantities 
(see the next section) 
the simpler case of open boundary conditions.
However, for the numerical techniques used in Sec.~\ref{sec5} to study finite systems, periodic boundary conditions are more appropriate, 
because more symmetries can be used, i.e., longer chains are feasible.

\section{Low-temperature thermodynamics}
\label{sec4}
\setcounter{equation}{0}

From previous investigations of model 3 it is known 
that the huge manifold of localized flat-band ground states may dominate the low-temperature thermodynamics
\cite{Zhitomirsky2004,Derzhko2004,Zhitomirsky2005,Derzhko2006,Richter2008}.
Since for models 1 and 2 the manifold of flat-band ground states is even significantly larger, 
this statement is certainly valid also for these sawtooth-chain spin models.
 
We consider the influence of a magnetic field $h$
on the manifold of the localized ground states of the open sawtooth-chain model 1 with $N=2{\cal{N}}+1$ sites.
For $h \ne 0$ only the single fully polarized ferromagnetic state remains the ground state with energy $E_0(h)=E_0-hN/2$, 
and all the other localized flat-band states become excited states. 
The contribution of all these states to the partition function is determined by 
their degeneracy $g_{\cal{N}}(S^z)$ given in Eq.~\eqref{312} 
and 
their Zeeman energy $E(h,S^z)=E_0-hS^z$:
\begin{eqnarray}
\label{401}
&&Z_{\rm{fbs}}(T,h,N)=2\exp\left(-\frac{E_0}{T}\right)
\nonumber\\
&& \quad 
\times\sum_{k=0}^{\cal{N}}
\left({\cal{C}}_{\cal{N}}^0+{\cal{C}}_{\cal{N}}^1+\ldots +{\cal{C}}_{\cal{N}}^k\right)
\cosh\frac{\left({\cal{N}}+\frac{1}{2}-k\right) h}{T}
\nonumber\\
&& \quad =\exp\left(-\frac{E_0}{T}\right)\sum_{k=0}^{\cal N} {\cal{C}}_{\cal{N}}^k F_k(x,{\cal N}),
\nonumber\\
&& F_k(x,{\cal N})=\frac{\sinh\left[({\cal{N}}+1-k)x\right]}{\sinh\frac{x}{2}},
\;\;\;
x=\frac{h}{T}.
\end{eqnarray}

As mentioned already above, this part of the partition function is identical for models 1 and 2 
(but not for model 3, where no complexes of overlapping localized magnons exist)
and may dominate the low-temperature physics.
It yields thermodynamic quantities which depend on $x=h/T$ only.
Clearly, the full partition functions of models 1 and 2 are different 
because of different excited non-flat-band states which come into play at nonzero temperatures,
and we can reveal these differences by an exact-diagonalization analysis of finite chains, see Sec.~\ref{sec5}.
We note that (how it should be) for $h=0$, $T\to 0$ the partition function reproduces the total ground-state degeneracy \eqref{313}, 
i.e., $Z_{\rm{fbs}}(x=0,N)={\cal{W}}(N)\exp(-E_0/T)$.

The residual ground-state entropy is given by $s=\ln [{\cal{W}}(N)]/N$, 
i.e., we get in the thermodynamic limit for the residual entropy per spin $s=\ln 2/2\approx 0.346\,574$.
As mentioned above, for $N \to \infty$  the boundary conditions become irrelevant: 
Because only the exponential term in Eqs.~\eqref{313} and \eqref{316} is essential, 
we get the same value for open and periodic chains.
Obviously, $s=s(T=0)$ is already half of the maximum entropy for $T \to \infty$. 
Note that the residual entropy for model 3 at the saturation field is smaller,
$s =(1/2)\ln[(1+\sqrt{5})/2] \approx 0.240\,606$ \cite{Zhitomirsky2004,Derzhko2004,footnote2}. 
Interestingly,
although the residual entropy following from Eq.~(\ref{313}) 
resembles that of flat-band systems of the so-called monomer universality class \cite{Derzhko2004,Derzhko2006},
the universal magneto-thermodynamics for both systems is different:
For monomer flat-band systems the partition function is as in Eq.~(\ref{401}), 
however, 
with $F_k(x,{\cal{N}})=\exp(kx)$,
see Ref.~\cite{Derzhko2006}.

A nonzero residual ground-state entropy leads to efficient magnetic cooling \cite{Honecker2004,Derzhko2006,Schnack2007}.
Importantly, for models 1 and 2 the residual entropy is present at zero field, 
i.e., it is relevant for cooling by varying the field around zero \cite{Evangelisti2010,Garlatti2013},
which is obviously an advantage from the practical point of view compared to model 3. 

From the Helmholtz free energy obtained from the partition function \eqref{401} by $F_{\rm{fbs}}(T,h,N)=-T\ln Z_{\rm{fbs}}(T,h,N)$ 
we get thermodynamic quantities such as magnetization, susceptibility, entropy or specific heat.
The magnetization $M=-\partial F/\partial h$ and the susceptibility $X=\partial M/\partial h$ are given by the formulas
\begin{eqnarray}
\label{402}
M_{\rm{fbs}}(x,N)
=\frac{\sum_{k=0}^{\cal N}{\cal{C}}_{\cal{N}}^k \frac{\partial F_k(x,{\cal{N}})}{\partial x}}{\sum_{k=0}^{\cal N}{\cal{C}}_{\cal{N}}^k F_k(x,{\cal{N}})}
\end{eqnarray}
and
\begin{eqnarray}
\label{403}
&& TX_{\rm{fbs}}(x,N) 
= 
\nonumber\\
&& \quad \frac{\sum_{k=0}^{\cal N}{\cal{C}}_{\cal{N}}^k \frac{\partial^2 F_k(x,{\cal{N}})}{\partial x^2}
\!-\!\left[\sum_{k=0}^{\cal N}{\cal{C}}_{\cal{N}}^k \frac{\partial F_k(x,{\cal{N}})}{\partial x}\right]^2}
{\left[\sum_{k=0}^{\cal N}{\cal{C}}_{\cal{N}}^k F_k(x,{\cal{N}})\right]^2},
\end{eqnarray}
respectively.
The entropy $S=-\partial F/\partial T$ and the specific heat $C=T\partial S/\partial T$ are given by the formulas
\begin{eqnarray}
\label{404}
S_{\rm{fbs}}(x,N)
&=&\ln\sum_{k=0}^{\cal N}{\cal{C}}_{\cal{N}}^k F_k(x,{\cal{N}})
-x\frac{\sum_{k=0}^{\cal N}{\cal{C}}_{\cal{N}}^k \frac{\partial F_k(x,{\cal{N}})}{\partial x}}{\sum_{k=0}^{\cal N}{\cal{C}}_{\cal{N}}^k F_k(x,{\cal{N}})}
\nonumber\\
&=&\ln Z_{\rm{fbs}}(x,N) - xM_{\rm{fbs}}(x,N)
\end{eqnarray}
and
\begin{eqnarray}
\label{405}
C_{\rm{fbs}}(x,N) 
&=& x^2
\frac{\sum_{k=0}^{\cal N}{\cal{C}}_{\cal{N}}^k \frac{\partial^2 F_k(x,{\cal{N}})}{\partial x^2}
\!-\!\left[\sum_{k=0}^{\cal N}{\cal{C}}_{\cal{N}}^k \frac{\partial F_k(x,{\cal{N}})}{\partial x}\right]^2}
{\left[\sum_{k=0}^{\cal N}{\cal{C}}_{\cal{N}}^k F_k(x,{\cal{N}})\right]^2} 
\nonumber\\
&=& x^2 TX_{\rm{fbs}}(x,N),
\end{eqnarray}
respectively.
Clearly,
the magnetization is an odd function of $x$,
whereas the susceptibility, the entropy, and the specific heat are even functions of $x$.
Below we consider the thermodynamic quantities per site to be denoted by small letters, e.g., $m=M/N$ etc.

\begin{figure}[ht!]
\centering
\includegraphics[clip=on,width=80mm,angle=0]{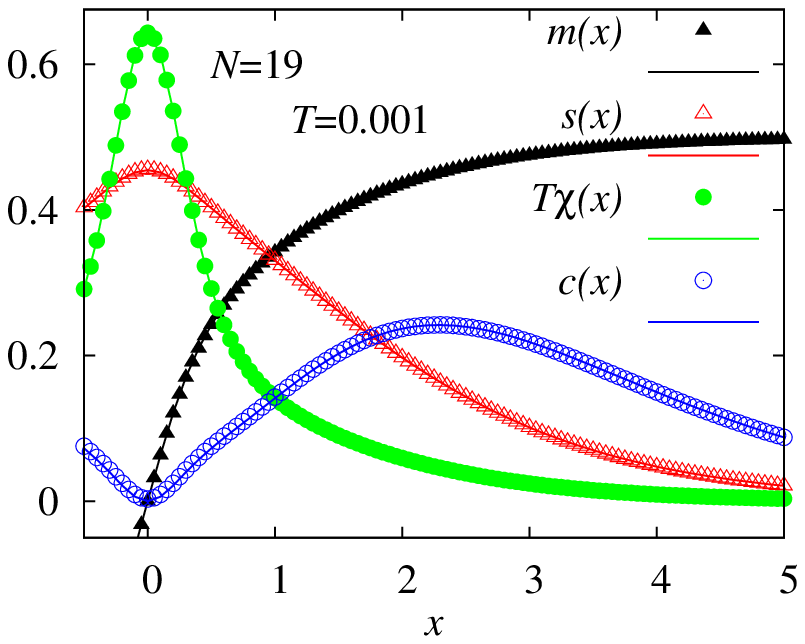}
\includegraphics[clip=on,width=80mm,angle=0]{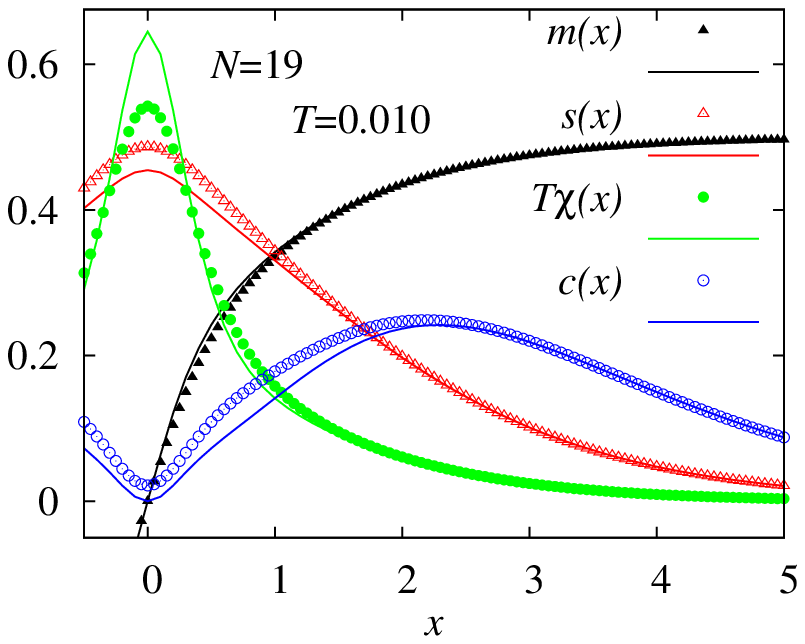}
\includegraphics[clip=on,width=80mm,angle=0]{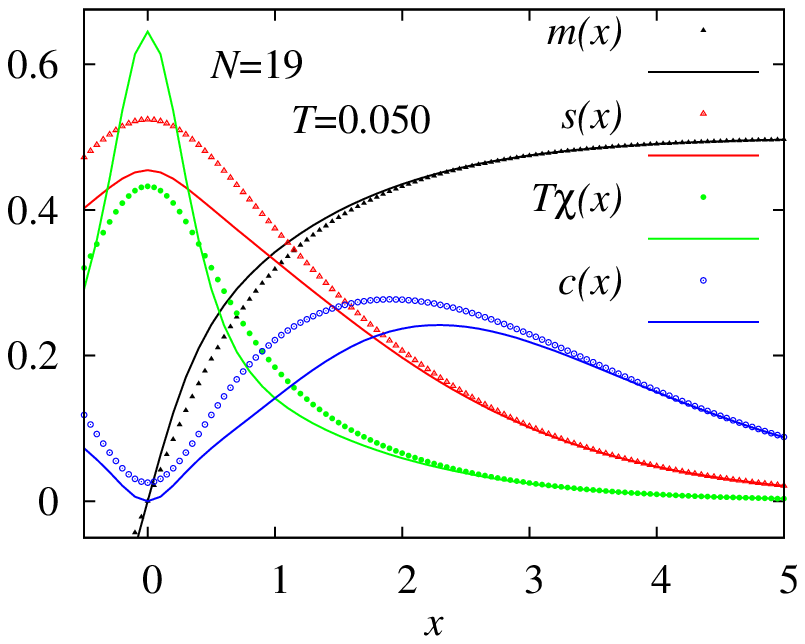}
\caption{Universal dependences $m(x)$, $s(x)$, $T\chi(x)$, and $c(x)$ versus $x=h/T$ 
for the open sawtooth-chain model 1 with $N=19$ (${\cal{N}}=9$). 
Symbols correspond to the full model ($J_1=1$) at different temperatures $T=0.001,\,0.01,\,0.05$ (from top to bottom), 
lines correspond to formulas (\ref{402}) -- (\ref{405}).} 
\label{f04} 
\end{figure}

The contribution of the flat-band states to the partition function 
is identical for models 1 and 2 in the thermodynamic limit as well as for finite open sawtooth chains. 
This contribution was discussed in detail in Ref.~\cite{Krivnov2014}.
In particular,
it was shown that the magnetization $m_{\rm{fbs}}(x)$ calculated in such a reduced basis depends essentially on $N$
and in the thermodynamic limit $N\to\infty$ it tends to 1/4 at $x = h/T \to 0$
in contradiction to the theorem that the magnetization should vanish in vanishing field at $T>0$ for one-dimensional systems. 
However, for finite chains the ``reduced-set'' magnetization given by Eq.~\eqref{402} may give a good estimate,
see Fig.~3 and discussion after Eq.~(31) in Ref.~\cite{Krivnov2014}.

\begin{figure}[ht!]
\begin{center}
\includegraphics[clip=on,width=80mm,angle=0]{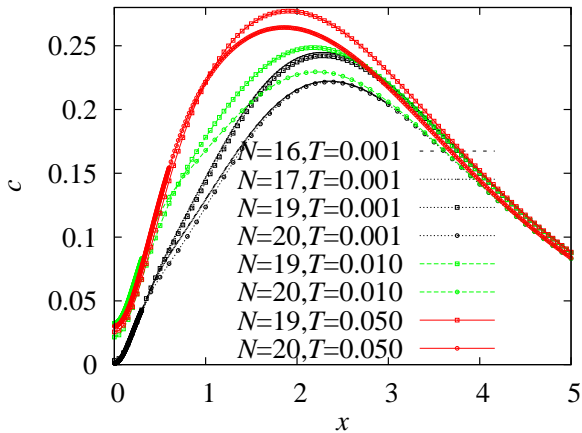}
\caption{Specific heat versus $x=h/T$ at low temperatures $T=0.001$ (black), $T=0.01$ (green), $T=0.05$ (red) 
for the sawtooth-chain model 1 ($J_1=1$) of $N=16,\,17,\,19,\,20$ sites.}
\label{f05}
\end{center}
\end{figure}

To illustrate the contribution of the localized flat-band states to the low-temperature thermodynamics in a magnetic field,
we compare in Fig.~\ref{f04} thermodynamic quantities as they follow from Eqs.~(\ref{402}) -- (\ref{405}) 
with exact-diagonalization data for the full open sawtooth-chain model 1 ($J_1=1$) with $N=19$.
While for $T=0.001$ the results are indistinguishable (top panel),
for $T=0.05$ the difference is definitely seen (bottom panel).
Nevertheless, 
the shape of the curves of the full model and of the universal behavior given by Eqs.~(\ref{402}) -- (\ref{405}) is still very similar at $T=0.05$. 
This comparison indicates the region of temperatures and fields 
within which the universal behavior (\ref{402}) -- (\ref{405}) determined by $x=h/T$ emerges.
It is obvious that for larger magnetic fields 
(i.e., large values of $x$ in Fig.~\ref{f04}) 
the localized-magnon states cover the thermodynamics not only at very low temperatures.

In Fig.~\ref{f05} we show the specific heat in dependence on $x=h/T$ at the same three temperatures as in Fig.~\ref{f04} 
for $N=17$ and $N=19$ (open boundary conditions imposed) and $N=16$ and $N=20$ (periodic boundary conditions imposed).
Comparing the data for $N=16$ and $N=20$ as well as for $N=17$ and $N=19$ at $T=0.001$ (black curves)
it is evident that the finite-size effects are small
(finite-size effects cannot be seen in this scale for $T=0.01$ and $T=0.05$ 
and therefore the data for $N=16,17$ are not shown at these temperatures).
However, comparing, e.g., $N=19$ with $N=20$
there is a noticeable influence of the boundary conditions for the considered finite values of $N$.

As already mentioned above, 
the universal behavior is identical for models 1 and 2 (but not for model 3).
Leaving the range of validity of formulas (\ref{402}) -- (\ref{405}) which display the contributions of flat-band states only,
the physics of the full spin model is determined more and more also by the non-flat-band states. 
This issue is studied in the next section by large-scale numerical calculations of the full spin models 1, 2 and 3.   

\section{Numerical investigations of finite systems}
\label{sec5}
\setcounter{equation}{0}

Now we consider models 1 and 2 at zero field and model 3 at the saturation field, 
i.e., all excitations are non-flat-band states.
We present full exact diagonalization (ED) and finite-temperature Lanczos (FTL) data 
for the considered sawtooth-chain models of finite size $N$. 
We focus here on periodic chains allowing to access larger system-sizes $N$ by exploiting the translational symmetry not present in open chains.   
Using ED we can study up to $N=20$ and using FTL we will provide data up to $N=32$. 

We begin with the excitation gaps $\Delta^{(i)}(S^z)$
($i=1$ corresponds to model 1 and $i=2$ corresponds to model 2), 
considered in each sector of $S^z$ separately, see Table~\ref{tab1} for $N=20$
(and also Table~I in Ref.~\cite{Krivnov2014} for model 2 with $N=16,20,24,28$ and $k=1,\ldots,6$).
For both models the gaps $\Delta^{(i)}(S^z=N/2-k)$ are rather small if $k>1$. 
However, for model 2 the gap $\Delta^{(2)}$ becomes virtually zero for $k>4$, but not for model 1.
This means that the contribution of low-lying excited states to the partition function enters for model 2 at lower temperatures than for model 1,
compare the specific heat in the low-temperature region shown in Fig.~\ref{f07} below.

\begin{table}
\begin{centering}
\caption{Ground-state degeneracies $G^{(i)}(S^z)$ and excitation gaps $\Delta^{(i)}(S^z)\equiv E^{(i)}_1(S^z)-E_0$
for the periodic sawtooth-chain model 1 ($J_1=1$, $i=1$) and model 2 ($J_1=1/2$, $i=2$) both of $N=20$ sites (${\cal{N}}=10$) 
in different subspaces $S^z$.
$E^{(i)}_1(S^z)$ is the energy of the lowest excitation in the subspace $S^z$ 
and 
$E_0=-3.75$ is the ground-state energy (which is identical for models 1 and 2). 
The results in the third and fifth columns coincide with the predictions according to Eqs.~\eqref{315} and \eqref{314}.
For (finite) periodic sawtooth chains of $N=20$ sites, 
the total number of ground states for model 1 is 4445 and for model 2 it is 3545.
\label{tab1}}
\vspace{3mm}
\begin{tabular}{|c|c||c|c||c|c|}
\hline 
$S^z$ & $k$ & $G^{(1)}(S^z)$ & $\Delta^{(1)}(S^z)$ & $G^{(2)}(S^z)$ & $\Delta^{(2)}(S^z)$ \tabularnewline
\hline 
\hline 
9  &  1 &  10                & 0.5                 &   10           & 1.0                 \tabularnewline
8  &  2 &  45                & 0.026\,996\,110\,0  &   45           & 0.021\,776\,745\,4  \tabularnewline
7  &  3 & 121                & 0.011\,213\,200\,0  &  120           & 0.000\,484\,876\,3  \tabularnewline
6  &  4 & 220                & 0.005\,858\,780\,0  &  210           & 0.000\,013\,213\,8  \tabularnewline
5  &  5 & 297                & 0.002\,110\,250\,0  &  252           & 0.000\,000\,197\,4  \tabularnewline
4  &  6 & 332                & 0.002\,256\,090\,0  &  252           & 0.000\,000\,064\,1  \tabularnewline
3  &  7 & 341                & 0.003\,116\,320\,0  &  252           & 0.000\,000\,064\,1  \tabularnewline
2  &  8 & 342                & 0.003\,828\,620\,0  &  252           & 0.000\,000\,035\,8  \tabularnewline
1  &  9 & 342                & 0.003\,247\,900\,0  &  252           & 0.000\,000\,007\,5  \tabularnewline
0  & 10 & 343                & 0.003\,792\,860\,0  &  253           & 0.000\,000\,007\,5  \tabularnewline
\hline
\end{tabular}
\par\end{centering}
\end{table}
  
\begin{figure}
\begin{center}
\includegraphics[clip=on,width=80mm,angle=0]{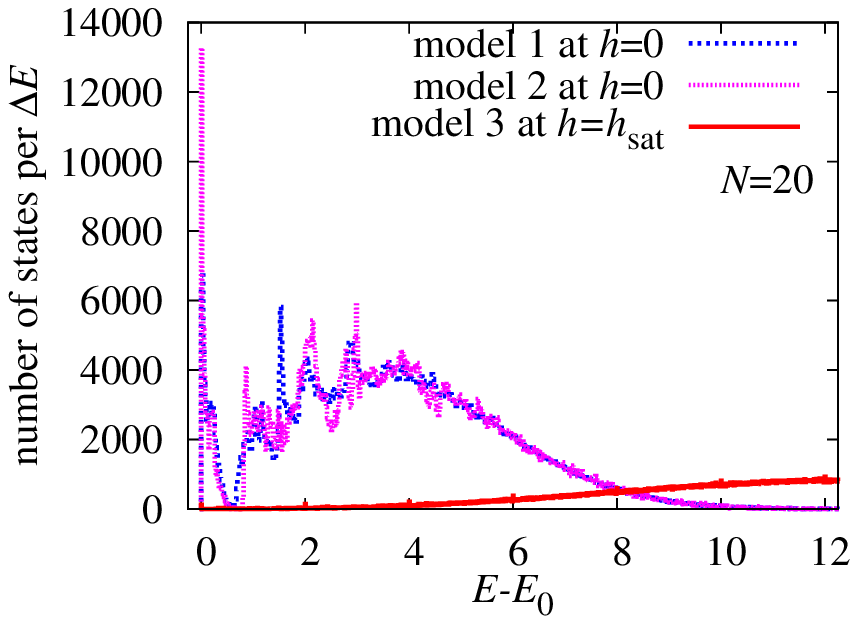}
\includegraphics[clip=on,width=80mm,angle=0]{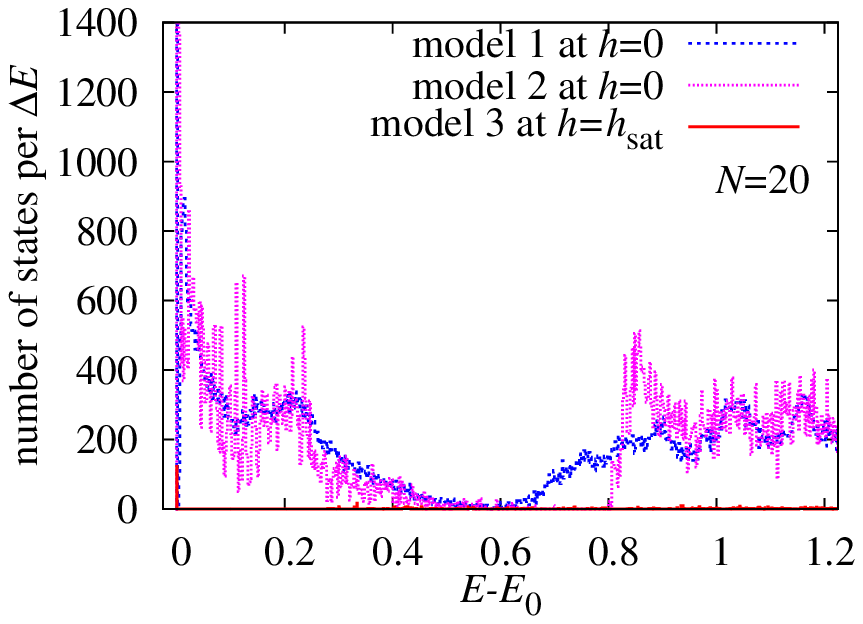}
\caption{Density of states (histogram) 
for the models 1 and 2 at $h=0$ (blue and magenta) and for model 3 at $h=h_{\rm sat}$ (red) 
for periodic chains of $N=20$ sites (ED).
Top: Histogram bar width $\Delta E=0.02$.
Bottom: Histogram bar width $\Delta E=0.002$, only low-energy part, where the $y$-axis is cut at $1400$ to improve the visibility.}
\label{f06}
\end{center}
\end{figure}

Next, we pass to the density of states, see Fig.~\ref{f06}.
Although the models 1 ($J_1=1$, blue curves) and 2 ($J_1=1/2$, magenta curves) do not have identical energy spectra, 
a similarity between both models is evident. 
On the other hand,   
the density of states of model 3 ($J_1=1/2$, red curves in Fig.~\ref{f06}) is completely different.
A striking feature of the density of states of models 1 and 2 
is the collection of about $6\%$ of the states in the low-energy region below $E-E_0 \lesssim 0.6$, 
where this region is separated by a quasi-gap from the high-energy region  $E-E_0 \gtrsim 0.6$.
This feature together with the huge ground-state degeneracy 
is responsible for the unconventional low-temperature physics of models 1 and 2.    

We have to comment on the height of the blue and magenta peaks at $E-E_0=0$ in Fig.~\ref{f06}.
As it as been explained above, 
the ground-state degeneracy for the periodic model 1 (4445) is larger than for the periodic model 2 (3545),
see Table~\ref{tab1}.
That would imply that the blue peak at $E-E_0=0$ is higher than the magenta one. 
However, the gaps for model 2 are much smaller than for model 1
and within the first histogram bar between $E_0$ and $E_0+\Delta E$ with $\Delta E=0.02$ or $\Delta E=0.002$ 
not only the ground states but also excited states are collected.
According to the above discussion of the gaps, 
there are a lot of excited states in the first $\Delta E$ interval for model 2 but much less for model 1. 

\begin{figure}
\begin{center}
\includegraphics[clip=on,width=80mm,angle=0]{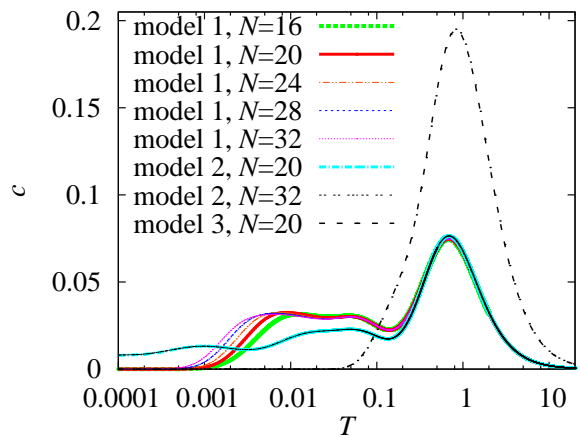}
\caption{Specific heat (logarithmic temperature scale) 
for periodic model 1 ($J_1=1$) and model 2 ($J_1=1/2$) at $h=0$ as well as for periodic model 3 ($J_1=1/2$) at $h=h_{\rm sat}$ ($h_{\rm sat}=2$).
Exact diagonalization data ($N=16,20$) 
and 
finite-temperature Lanczos data ($N=24,28,32$; $R=20$ for model 1 and $R=10$ for model 2).
Finite-size effects are small.}
\label{f07}
\end{center}
\end{figure}

\begin{figure}
\begin{center}
\includegraphics[clip=on,width=80mm,angle=0]{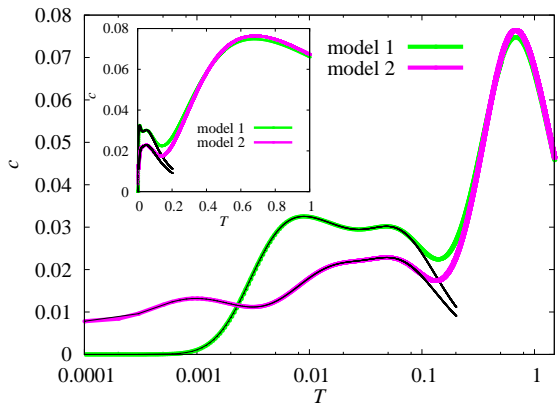}
\caption{Specific heat $c(T)$ (exact diagonalization data, periodic chain $N=20$) for model 1 ($J_1=1$) and model 2 ($J_1=1/2$) at $h=0$: 
Green and magenta lines -- exact data using the full spectrum,
black lines -- approximate data using only the low-energy spectrum below the quasi-gap.  
Main panel: Logarithmic temperature scale.  
Inset: Linear temperature scale.}
\label{f08}
\end{center}
\end{figure}

Finally,
in Fig.~\ref{f07} we show the temperature dependence (logarithmic temperature scale) of the specific heat for all three models 
(models 1 and 2 at zero field and model 3 at the saturation field). 
The specific heat of model 3 as characterized by single pronounced maximum
followed by an exponential decay of $c(T)$ as $T\to 0$ leading to a virtually vanishing  specific heat below $T\sim 0.06$.
By contrast, 
the very specific low-energy density of states of models 1 and 2 with much smaller energy gaps and the quasi-gap at about $E=E_0+0.06$
leads to a distinct separation of temperature scales in the temperature dependence of the specific heat  
which is characterized by a pronounced low-temperature profile of $c(T)$ with two additional maxima below the typical main maximum.
The difference in the details of the low-energy spectrum of the two models
results in a deviation of the $c(T)$ curves of both models at low temperatures starting at about $T=0.3$.
For model 1 the finite-size effects are negligible down to $T\sim 0.01$, 
whereas model 2 exhibits practically no finite-size effects in the temperature region shown in Fig.~\ref{f07}. 
This difference can be attributed to the different sizes of the excitation gaps, cf. Table~\ref{tab1}.

To demonstrate the relation of the separation of temperature scales in the $c(T)$ profile 
to the very specific structure of the density of states of the models 1 and 2 
we show in Fig.~\ref{f08} the specific heat $c(T)$  for periodic chains of $N=20$ sites using the full spectrum 
(i.e., numerically exact data)
together with the approximate data for $c(T)$ which are calculated using a restricted set of energies $E < E_0+0.06$,  
i.e., only the low-energy spectrum below the quasi-gap is taken into account.  
This comparison reveals 
that indeed the unconventional features in $c(T)$ below the main maximum are entirely covered by the energy levels below the quasi-gap.

\section{Conclusions}
\label{sec6}
\setcounter{equation}{0}

To summarize,
in the present paper we have examined three spin-1/2 sawtooth-chain models 
which for a special choice of parameters exhibit flat-band physics.
While the universal flat-band behavior of the two models introduced in Ref.~\cite{Changlani2018} and in Ref.~\cite{Krivnov2014} is identical,
it is different from the universal flat-band behavior of the model introduced in Refs.~\cite{Schulenburg2002,Derzhko2006}:
In the latter case flat-band many-magnon localized states are independent localized magnons only,
whereas in the former two cases the flat-band ground-state manifold is larger 
containing in addition specifically overlapping localized magnons (many-magnon complexes).
Flat-band states dominate the low-temperature thermodynamics around zero field (models 1 and 2) or around the saturation field (model 3).
The localized nature of flat-band states allows the complete analysis of the massively degenerate ground-state manifold, 
and, as a result, to find analytical expressions for the low-temperature thermodynamics in presence of a magnetic field.
These analytical expressions are complemented by numerical calculations of thermodynamic quantities of finite sawtooth spin chains.         

The spin-1/2 $XXZ0$ sawtooth-chain model admits another promising route 
to examine its properties using the three-coloring representation \cite{Changlani2018,Changlani2019}.
Interesting and still open problems are:
How to present the localized many-magnon complex states within the three-coloring picture?
Is it possible to use the three-coloring representation for constructing thermodynamics?
In the future,
it might be interesting to consider small perturbations to the basic flat-band sawtooth-chain models (\ref{101}).
One extension is to introduce into Hamiltonian (\ref{101}) the interaction between neighboring apical sites $J_1^\prime$, $\Delta_1^\prime$ 
to see traces of the flat-band-states manifolds.  

\section*{Acknowledgments}

O.~D. acknowledges the kind hospitality of the MPIPKS, Dresden in September-November of 2019.
This work was supported by the Deutsche Forschungsgemeinschaft DFG
(314331397 (SCHN 615/23-1); 355031190 (FOR~2692); 397300368 (SCHN~615/25-1)). 
Computing time at the Leibniz Center in Garching is gratefully acknowledged.

\appendix
\section{Some analytical calculations supplementary to Sec.~\ref{sec3}}
\label{appendix}
\renewcommand{\theequation}{A.\arabic{equation}}
\setcounter{equation}{0}

Here we present some detailed analytical calculations mentioned but not presented in the main text.

First we check 
that in the subspace $S^z=N/2-1$ 
the boundary and bulk localized-magnon states defined in Eqs.~(\ref{302}) and (\ref{303}) 
are exact ground states for model 1 of $N=2{\cal N}+1$ sites with open boundary conditions,  
where ${\cal{N}}$ is the number of triangles in this chain.
We begin with the boundary state 
\begin{eqnarray}
\label{a01}
\vert l_1\rangle
= 
\left(-\vert\downarrow_1\uparrow_2\uparrow_3\rangle + \vert\uparrow_1\downarrow_2\uparrow_3\rangle\right)\vert\ldots\uparrow\ldots\rangle.
\end{eqnarray}
That is a state with one flipped spin located on the boundary bond $J_2$ of the zig-zag path,
cf. Fig.~\ref{f01}, top.
This state contains the eigenstate $\vert2\rangle$ of the Hamiltonian of the first triangle $H_{123}$ with the energy $-3/8$, 
cf. Eq.~(\ref{211}),
and the Hamiltonians of all other triangles act on the fully polarized state $\vert 1\rangle$ (\ref{209}) giving each time again the energy $-3/8$.
Therefore, 
the state $\vert l_1\rangle$ (\ref{a01}) belongs to the ground-state manifold of model 1 with the energy $(-3/8){\cal{N}}$.
We pass to the bulk state
\begin{eqnarray}
\label{a02}
\vert l_2\rangle
= 
\left(\vert\downarrow_2\uparrow_3\uparrow_4\rangle-\vert\uparrow_2\downarrow_3\uparrow_4\rangle + \vert\uparrow_2\uparrow_3\downarrow_4\rangle\right)
\vert\ldots\uparrow\ldots\rangle, \qquad
\end{eqnarray}
where the numbers at the up- and down-arrows correspond to the numbering in Fig.~\ref{f01}, top.
We consider the application of $H_{123}$ and $H_{345}$ on $\vert l_2\rangle$ (\ref{a02})
(all Hamiltonians $H_\triangle$ of other triangles act on the fully polarized state $\vert 1\rangle_\triangle$ yielding the energy $-3/8$).
We notice that $H_{123}$ acts 
either on 
$\left(\vert\uparrow_1\downarrow_2\uparrow_3\rangle-\vert\uparrow_1\uparrow_2\downarrow_3\rangle\right)\vert\ldots\uparrow\ldots\rangle
=\left(\vert 3\rangle-\vert 2\rangle\right)_{123}\vert\ldots\uparrow\ldots\rangle$
or on
$\vert\uparrow_1\uparrow_2\uparrow_3\rangle\vert\ldots\rangle
=\vert 1\rangle_{123} \vert\ldots\rangle$,
i.e., on the eigenstates with the energy $-3/8$,
see Eqs.~(\ref{211}), (\ref{209}) in Sec.~\ref{sec2b}.
Furthermore,
$H_{345}$ acts 
either on
$\vert\uparrow_3\uparrow_4\uparrow_5\rangle\vert\ldots\rangle
=\vert 1\rangle_{345} \vert\ldots\rangle$
or on
$\left(-\vert\downarrow_3\uparrow_4\uparrow_5\rangle+\vert\uparrow_3\downarrow_4\uparrow_5\rangle\right)\vert\ldots\uparrow\ldots\rangle
=\left(-\vert 2\rangle\right)_{345}\vert\ldots\uparrow\ldots\rangle$,
i.e., once more on the eigenstates with the energy $-3/8$.
Hence, the state $\vert l_2\rangle$ (\ref{a02}) also belongs to the ground-state manifold of model 1.

Next we check that overlapping localized two-magnon complexes, 
Eqs.~(\ref{305}) and (\ref{306}), 
are eigenstates of the Hamiltonian (\ref{201}) with the eigenvalue $(-3/8){\cal N}$.
We start with the state (\ref{305}) 
and consider first how the Hamiltonian $H_{123}$ acts on this state.
It acts 
either on
$\left(-s_1^-+s_2^-\right)\left(s_1^-/2+s_2^-/2-s_3^-\right)\vert 0\rangle=\left(-\vert 5\rangle\right)_{123}\vert\ldots\uparrow\ldots\rangle$
or on 
$\left(-s_1^-+s_2^-\right)s_4^-\vert 0\rangle=\left(-\vert 2\rangle\right)_{123}\vert\ldots\rangle$.
In both cases,
$H_{123}$ acts on its eigenstates with the energy $-3/8$.
Next,
$H_{345}$ acts
either on
$\vert\uparrow_3\uparrow_4\uparrow_5\rangle\vert\ldots\rangle=\vert 1\rangle_{345}\vert\ldots\rangle$
or on
$\left(-s_3^-+s_4^-\right)\ldots\vert 0\rangle=\left(-\vert 2\rangle\right)_{345}\vert\ldots\rangle$,
i.e., on its eigenstates with the energy $-3/8$.
Finally,
all other $H_\triangle$ acts on
$\vert 1\rangle_{\triangle}\vert\ldots\rangle$,
i.e., on their eigenstates with the energy $-3/8$.
As a result, we conclude 
that the localized two-magnon complex $l_1\left(c l_{1} + l_{2}\right)\vert 0\rangle$ (\ref{305}) is indeed among the ground states with $S^z=N/2-2$.

Now we pass to the state (\ref{306}).
We have to consider the application of $H_{123}$, $H_{345}$, and $H_{567}$ on the state (\ref{306}),
since the rest Hamiltonians $H_\triangle$, 
while acting on the state (\ref{306}),
``see'' the only relevant factor $\vert 1\rangle_\triangle$ which is the eigenstate \eqref{209} with the eigenvalue $-3/8$.
We begin with $H_{123}$.
It acts 
either on 
$\left(s_2^--s_3^-\right)\left(-s_1^-+s_2^-/2+s_3^-/2\right)\vert 0\rangle
=\left(\vert 6\rangle-\vert 5\rangle\right)_{123}\vert\ldots\uparrow\ldots\rangle$,
or on
$\left(s_2^--s_3^-\right)\ldots\vert 0\rangle
=\left(\vert 3\rangle - \vert 2 \rangle\right)_{123}\vert\ldots\rangle$,
or on
$\left(-s_1^-+s_2^-/2+s_3^-/2\right)\ldots\vert 0\rangle
=\left(-\vert 2\rangle/2 - \vert 3 \rangle/2\right)_{123}\vert\ldots\rangle$,
or on
$\vert 1\rangle_{123}\vert\ldots\rangle$,
i.e., each time on the eigenstates with the eigenvalue $-3/8$.
Next,
$H_{345}$ acts 
either on 
$\vert 1\rangle_{345}\vert\ldots\rangle$,
or on
$\left(s_3^-/2+s_4^-/2-s_5^-\right)\ldots\vert 0\rangle
=\left(\vert 3\rangle - \vert 2 \rangle/2\right)_{123}\vert\ldots\rangle$,
or on
$\left(-s_3^-+s_4^-\right)\ldots\vert 0\rangle
=\left(- \vert 2 \rangle\right)_{345}\vert\ldots\rangle$,
or on
$\left(-s_3^-+s_4^-\right)\left(s_3^-/2+s_4^-/2-s_5^-\right)\vert 0\rangle
=\left(-\vert 5\rangle\right)_{345}\vert\ldots\uparrow\ldots\rangle$,
i.e., each time on the eigenstates with the eigenvalue $-3/8$.
Finally, 
$H_{567}$ acts 
either on 
$\vert 1\rangle_{567}\vert\ldots\rangle$
or on
$\left(-s_5^-+s_6^-\right)\ldots\vert 0\rangle
=\left(- \vert 2 \rangle\right)_{567}\vert\ldots\rangle$,
i.e., each time on the eigenstates with the eigenvalue $-3/8$.
In sum,
the state 
$l_2\left(l_{1}+c l_{2} + l_{3}\right)\vert 0\rangle$ (\ref{306})
belongs to the ground-state manifold with $S^z=N/2-2$.
Similar calculations for the states 
$l_3\left(l_{2}+cl_{3} + l_{4}\right)\vert 0\rangle$,
\ldots,
$l_{{\cal N}+1}\left(l_{\cal{N}}+cl_{{\cal N}+1}\right)\vert 0\rangle$
confirm that this kind of states belongs to the ground-state manifold with $S^z=N/2-2$.

We consider now a localized three-magnon complex of the one-bracket type given in Eq.~(\ref{308}).
In more detail it reads
\begin{eqnarray}
\label{a03}
&& l_2l_4\left(l_{1}+cl_{2} + l_{3} +cl_{4} + l_{5}\right)\vert 0\rangle
\\
&&=\left(s_2^--s_3^-+s_4^-\right)\left(s_6^--s_7^-+s_8^-\right) \!\times\!
\nonumber\\
&&
\left(\!-s_1^-\!+\!\frac{s_2^-}{2}\!+\!\frac{s_3^-}{2}\!+\!\frac{s_4^-}{2}\!-\!s_5^-
\!+\!\frac{s_6^-}{2}\!+\!\frac{s_7^-}{2}\!+\!\frac{s_8^-}{2}\!-\!s_9^-\!+\!s_{10}^-\!\right)
\!\vert 0\rangle \qquad \nonumber
\end{eqnarray}
($c=-1/2$),
see line 3 in Fig.~\ref{f03}.
We have to check 
whether this state is an eigenstate of the Hamiltonians $H_{123}$, $H_{345}$, $H_{567}$, $H_{789}$, and $H_{9,10,11}$ with the eigenvalue $-3/8$.
As explained above,
the Hamiltonians $H_{123}$ and $H_{567}$ while acting on the state \eqref{a03} ``see'' only their eigenstates
$\vert 6\rangle - \vert 5\rangle$,
$\vert 3\rangle - \vert 2\rangle$,
$-\vert 2\rangle/2 - \vert 3\rangle/2$,
and 
$\vert 1\rangle$;
all with the eigenvalue $-3/8$.
Next,
the Hamiltonians $H_{345}$ and $H_{789}$ while acting on the state \eqref{a03} ``see'' only their eigenstates
$- \vert 5\rangle$,
$- \vert 2\rangle$,
$- \vert 2\rangle/2 + \vert 3\rangle$,
and 
$\vert 1\rangle$;
all with the eigenvalue $-3/8$.
Finally,
and the Hamiltonian $H_{9,10,11}$ while acting on the state \eqref{a03} ``sees'' only its eigenstates
$-\vert 2\rangle_{9,10,11}$
and 
$\vert 1\rangle_{9,10,11}$;
both with the eigenvalue $-3/8$.
Hence, the state given in Eq.~\eqref{a03} [or Eq.~\eqref{308}] belongs to the ground-state manifold with $S^z=N/2-3$.

Now we check that localized three-magnon complexes of the two-bracket type are ground states.
We consider, for example, the state given in the third line of Eq.~\eqref{310},
i.e.,
\begin{eqnarray}
\label{a04}
\left[
\left(\!s^-_4-s^-_5+s^-_6\!\right) 
\!
\left(\!s^-_2-s^-_3+\frac{s^-_4}{2}+\frac{s^-_5}{2}+\frac{s^-_6}{2}-s^-_7+s^-_8\!\right)
\right.
\nonumber\\
\left.
\times
\!
\left(\!-s^-_1\!+\!\frac{s^-_2}{2}\!+\!\frac{s^-_3}{2}\!-\!\frac{s^-_4}{4}\!-\!\frac{s^-_5}{4}\!-\!\frac{s^-_6}{4}
\!+\!\frac{s^-_7}{2}\!+\!\frac{s^-_8}{2}\!-\!s^-_9\!+\!s^-_{10}\!\right)
\right.
\nonumber\\
\left.
-\frac{3}{4}s^-_4s^-_5s^-_6
\right]\vert 0\rangle, \qquad \qquad
\end{eqnarray}
see line 6 in Fig.~\ref{f03}.
Checking that the Hamiltonians $H_{123}$, $H_{789}$, $H_{9,10,11}$ and so on while acting on the state \eqref{a04} give $-3/8$ multiplied by this state 
is straightforward by repetition of the calculations explained above.
The role of the terms with $d$ in Eq.~\eqref{310} becomes clear
after acting on the state \eqref{a04} by the Hamiltonians $H_{345}$ and $H_{567}$:
Only after accounting the term $(-3/4)s_4^-s_5^-s_6^-$
these Hamiltonians ``see'' some linear combinations of their eigenstates with the eigenvalue $-3/8$.
As a result,
in all cases we arrive at the state \eqref{a04} multiplied by $-3/8$.
Hence the state \eqref{a04} is within the ground-state manifold with $S^z=N/2-3$.


\begin{thebibliography}{99}

\bibitem{Derzhko2015}
O.~Derzhko, J.~Richter, and M.~Maksymenko,
Int. J. Mod. Phys. B {\bf 29}, 1530007 (2015).

\bibitem{Zhitomirsky2004}
M.~E.~Zhitomirsky and H.~Tsunetsugu,
Phys. Rev. B {\bf 70}, 100403(R) (2004).

\bibitem{Derzhko2004}
O.~Derzhko and J.~Richter,
Phys. Rev. B {\bf 70}, 104415 (2004).

\bibitem{Zhitomirsky2005}
M.~E.~Zhitomirsky and H.~Tsunetsugu,
Progress of Theoretical Physics Supplement 160, 361 (2005).

\bibitem{Derzhko2006}
O.~Derzhko and J.~Richter,
Eur. Phys. J. B {\bf 52}, 23 (2006).

\bibitem{Zhitomirsky2007}
M.~E.~Zhitomirsky and H.~Tsunetsugu,
Phys. Rev. B {\bf 75}, 224416 (2007).

\bibitem{Schnack2018}
J.~Schnack, J.~Schulenburg, and J.~Richter,
Phys. Rev. B {\bf 98}, 094423 (2018).

\bibitem{Mielke1991}
A.~Mielke, 
J. Phys. A {\bf 24}, L73 (1991);
A.~Mielke, 
J. Phys. A {\bf 24}, 3311 (1991).

\bibitem{Tasaki1992}
H.~Tasaki,
Phys. Rev. Lett. {\bf 69}, 1608 (1992).

\bibitem{Derzhko2010}
O.~Derzhko, J.~Richter, A.~Honecker, M.~Maksymenko, and R.~Moessner,
Phys. Rev. B {\bf 81}, 014421 (2010).

\bibitem{Maksymenko2012}
M.~Maksymenko, A.~Honecker, R.~Moessner, J.~Richter, and O.~Derzhko,
Phys. Rev. Lett. {\bf 109}, 096404 (2012).

\bibitem{Schulenburg2002}
J.~Schulenburg, A.~Honecker, J.~Schnack, J.~Richter, and H.-J.~Schmidt,
Phys. Rev. Lett. {\bf 88}, 167207 (2002).

\bibitem{Richter2008}
J.~Richter, O.~Derzhko, and A.~Honecker,
Int. J. Mod. Phys. B {\bf 22}, 4418 (2008).

\bibitem{Metavitsiadis2020}
A.~Metavitsiadis, C.~Psaroudaki, and W.~Brenig,
arXiv:2002.07190.

\bibitem{Krivnov2014}
V.~Ya.~Krivnov, D.~V.~Dmitriev, S.~Nishimoto, S.-L.~Drechsler, and J.~Richter,
Phys. Rev. B {\bf 90}, 014441 (2014).

\bibitem{Tonegawa2004} 
T.~Tonegawa and M.~Kaburagi, 
J. Magn. Magn. Mater. {\bf 272-276}, 898 (2004).

\bibitem{Kaburagi2005} 
M.~Kaburagi, T.~Tonegawa, and M.~Kang, 
J. Appl. Phys. {\bf 97}, 10B306 (2005).

\bibitem{Dmitriev2015}
D.~V.~Dmitriev and V.~Ya.~Krivnov,
Phys. Rev. B {\bf 92}, 184422 (2015).

\bibitem{Dmitriev2017}
D.~V.~Dmitriev and V.~Ya.~Krivnov,
J. Phys.: Condens. Matter {\bf 29}, 215801 (2017).

\bibitem{Dmitriev2019}
D.~V.~Dmitriev, V.~Ya.~Krivnov, J.~Richter, and J.~Schnack,
Phys. Rev. B {\bf 99}, 094410 (2019).

\bibitem{Dmitriev2020}
D.~V.~Dmitriev, V.~Ya.~Krivnov, J.~Schnack, and J.~Richter,
Phys. Rev. B {\bf 101}, 054427 (2020).

\bibitem{Baniodeh2018}
A.~Baniodeh, N.~Magnani, Y.~Lan, G.~Buth, C.~E.~Anson, J.~Richter, M.~Affronte, J.~Schnack, and A.~K.~Powell, 
npj Quantum Materials {\bf 3}, 10 (2018).

\bibitem{Shirakami2019}
R.~Shirakami, H.~Ueda, H.~O.~Jeschke, H.~Nakano, S.~Kobayashi, A.~Matsuo, T.~Sakai, N.~Katayama, H.~Sawa, K.~Kindo, C.~Michioka, and K.~Yoshimura, 
Phys. Rev. B {\bf 100}, 174401 (2019).

\bibitem{Changlani2018}
H.~J.~Changlani, D.~Kochkov, K.~Kumar, B.~K.~Clark, and E.~Fradkin,
Phys. Rev. Lett. {\bf 120}, 117202 (2018).

\bibitem{Changlani2019}
H.~J.~Changlani, S.~Pujari, C.-M.~Chung, and B.~K.~Clark,
Phys. Rev. B {\bf 99}, 104433 (2019).

\bibitem{Luban2001}
M.~Axenovich and M.~Luban,
Phys. Rev. B {\bf 63}, 100407(R) (2001).

\bibitem{Cepas2011}
O.~Cepas and A.~Ralko,
Phys. Rev. B {\bf 84}, 020413(R) (2011).

\bibitem{Jaubert2016} 
K.~Essafi, O.~Benton, and L.~D.~C.~Jaubert, 
Nat. Commun. {\bf 7}, 10297 (2016).

\bibitem{spinpack} 
J.~Richter and J.~Schulenburg, 
Eur. Phys. J. B {\bf 73}, 117 (2010);
\newline
https://www-e.uni-magdeburg.de/jschulen/spinack

\bibitem{FTL1} 
J.~Jakli\v{c} and P.~Prelov\v{s}ek, 
Phys. Rev. B {\bf 49}, 5065(R) (1994).

\bibitem{FTL2}
J.~Jakli\v{c} and P.~Prelov\v{s}ek,
Adv. Phys. {\bf 49}, 1 (2000).

\bibitem{FTL3}
B.~Schmidt and P.~Thalmeier, 
Phys. Rep. {\bf 703}, 1 (2017).

\bibitem{FTL4} 
J.~Schnack and O.~Wendland, 
Eur. Phys. J. B {\bf 78}, 535 (2010); 
J.~Schnack, J.~Richter, and R.~Steinigeweg,
Phys. Rev. Research {\bf 2}, 013186 (2020).

\bibitem{footnote1}
The analogue of the state (\ref{302}) for the model 2 is
$\vert l_1\rangle
= 
\left(2\vert\downarrow_1\uparrow_2\uparrow_3\rangle + \vert\uparrow_1\downarrow_2\uparrow_3\rangle\right)\vert\ldots\uparrow\ldots\rangle$.

\bibitem{Schmidt2006}
H.-J.~Schmidt, J.~Richter, and R.~Moessner,
J. Phys. A {\bf 39}, 10673 (2006).   

\bibitem{footnote2}
This result follows from the formula 
for the total number of ground states for the open sawtooth chain 3 of $N=2{\cal{N}}+1$ sites at $h=h_{\rm{sat}}$,
\begin{eqnarray}
\nonumber
{\cal{W}}(N)=\sum_{k=0}^{\left[\frac{{\cal{N}}}{2}\right]} {\cal{C}}_{{\cal{N}}-k}^{k},
\end{eqnarray}
which implies the following recurrence relation:
\begin{eqnarray}
\nonumber
{\cal{W}}(N-4)+{\cal{W}}(N-2)={\cal{W}}(N)
\end{eqnarray}
(to prove it, one has to use the identity ${\cal{C}}_{{\cal{M}}}^{p}+{\cal{C}}_{{\cal{M}}}^{p+1}={\cal{C}}_{{\cal{M}}+1}^{p+1}$)
or
\begin{eqnarray}
\nonumber
\frac{{\cal{W}}(N-4)}{{\cal{W}}(N-2)} + 1  - \frac{{\cal{W}}(N)}{{\cal{W}}(N-2)}= 0.
\end{eqnarray}
Evidently, we have arrived at a Fibonacci sequence.
In the limit $N\to\infty$,
the ratio ${\cal{W}}(N-2) / {\cal{W}}(N) $ tends to $\varphi=(\sqrt{5}-1)/2$,
i.e.,
in the thermodynamic limit, 
${\cal{W}}(N)=\varphi^{-\frac{N}{2}}$ 
resulting in $s=(1/2)\ln[(1+\sqrt{5})/2]$.

\bibitem{Honecker2004} 
M.~E.~Zhitomirsky and A.~Honecker, 
J. Stat. Mech.: Theory Exp. P07012 (2004).

\bibitem{Schnack2007} 
J.~Schnack,  R.~Schmidt, and J.~Richter,
Phys. Rev. B {\bf 76}, 054413 (2007).

\bibitem{Evangelisti2010}
M.~Evangelisti and E.~K.~Brechin, 
Dalton Trans. {\bf 39}, 4672 (2010).

\bibitem{Garlatti2013}
E.~Garlatti, S.~Carretta, J.~Schnack, G.~Amoretti, and P.~Santini,
Appl. Phys. Lett. {\bf 103}, 202410 (2013).

\end{thebibliography}
\end{document}